\documentstyle[editedvolume,psfig]{crckapb}

\def\p{$\pm$}
\def\fdu{$\phi_{21}$}
\def\ftu{$\phi_{31}$}

\def\fu{$f_1$}
\def\fd{$f_2$}
\def\fp{$f_1 + f_2$}
\def\fm{$f_2 - f_1$}
\def\cd{cd$^{-1}$}
\def\cds{cd$^{-1}$\,}

\begin{opening}
\title{Asteroseismology of Cepheids}

\author{Ennio PORETTI}
\institute{Osservatorio Astronomico di Brera\\
Via E. Bianchi, 46 -- 23807 Merate, Italy}

\end{opening}

\runningtitle{Cepheids}

\begin{document}

\section{Introduction}
The importance of Cepheids is well known in many fields of astronomy.
In this contribution I  would like to show how it is possible to obtain
indications about the internal structure of a Cepheid and how we can test models
of this class of variable stars.

Section 2 introduces the Fourier decomposition,  a tool to
describe quantitatively the light curves of pulsating stars.
In the past years I have been involved in a project concerning Cepheids with
$P<$ 8 d and in Sect. 3 I  will show how an observational result was 
progressively built on the basis of old and new data; the latter
were collected on selected targets just to clarify some controversial
points. Since different pulsation modes were suspected among these stars,
an independent confirmation was searched for.

To do that we applied the least--squares technique to Double Mode 
Cepheids. By obtaining a very satisfactory description of their
pulsational content (Sect.~4), 
we demonstrated how powerful the method is. Moreover,
we could confirm that between Cepheids with
$P<$~8~d they are both fundamental and first overtone pulsators
(Sect.~5).  The detection of small amplitude cross--coupling terms and higher
harmonics in the light curves of Double Mode Cepheids allowed us to
quantify the properties of the high--order terms and hence to discover
other peculiarities (Sect.~6), very useful to complete the scenario of
the resonance effects  and to test some theoretical models (Sect.~7). 

\section{The application of Fourier decomposition to Cepheid light
curves}
Following the notation proposed by Simon and Lee (1981), the Fourier
decomposition  consists in interpolating the measurements by means of
the series 
\begin{equation}
     V(t)= A_o + \sum_{i=1}^N {A_i \cos [2\pi i(t-T_o)f +\phi_i ]}
\end{equation}
$V(t)$ is the magnitude observed at times $t$, $A_0$ the mean
magnitude, $A_i$ the amplitudes of each component, $f$ the frequency
($f$=1/$P$, where $P$ is the period of the light variation), $\phi_i$ the
$i$--th phase at $t=T_o$. The components $2f, 3f, 4f$~...~are the first, second,
third~....~harmonics of the main frequency $f$.
 Note that the use of
the {\tt sin}  term instead of the {\tt cos} one can lead to spurious results,
owing to the $\pi/2$ shift of the phase component. Another trouble 
can originate from a different formula, for example considering
$2\pi\phi_i$ in the development. 
 
This technique provides quantitative parameters to define the shape of the
light curves and it is therefore a powerful tool for classification purposes.
The Fourier parameters can be subdived into two groups: the amplitude
ratios $R_{ij}=A_i/A_j$ (i.e. $R_{21}=A_2/A_1$,  $R_{31}=A_3/A_1$,
$R_{32}=A_3/A_2$, ...) and the phase shifts $\phi_{ji}=i\phi_j-j\phi_i$
 (i.e. \fdu=$\phi_2-2\phi_1$,
\ftu=$\phi_3-3\phi_1$, $\phi_{32}=2\phi_3-3\phi_2$, ...).

Figure 1 shows how the light curve of a pulsating stars is
progressively changed by adding harmonic terms. The upper panel
represents a perfect sine--shaped curve having frequency $f$. When adding 
the first harmonic $2f$ ($R_{21}=$0.30 and
\fdu=4.5 rad), the light curve immediately becomes asymmetrical (middle
panel). Adding the second harmonic $2f$ ($R_{31}$=0.10 and \ftu=2.5 rad),
the brightness increase is again much steeper; the effect of higher
harmonics is  to originate curves which are even again more asymmetrical (i.e.
with a decreasing  
$M-m$ value, where $M$ is the phase of maximum brightness and $m$ the
phase of minimum brightness) and to fit some small jumps of the light
curves. The case of S Cru ($P$=4.68997 d, 6$^{\rm th}$ order fit) is shown
in Fig. 2.

\begin{figure}
\centerline{\psfig{figure=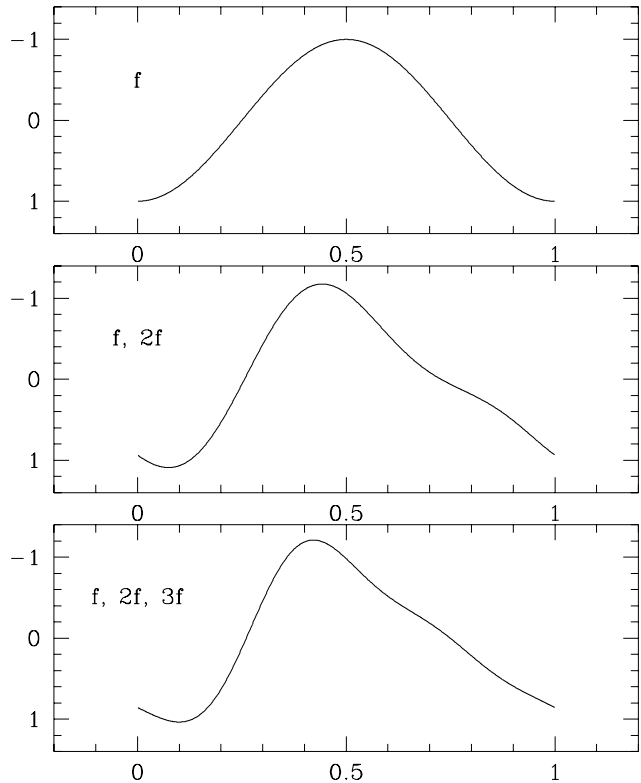,width=9truecm}}
\caption{Changes in the light curve by adding to a sine--shaped term
(upper panel) its first harmonic $2f$ (medium panel) and its second harmonic
$3f$ (lower panel)}
\end{figure}

\begin{figure}
\centerline{\psfig{figure=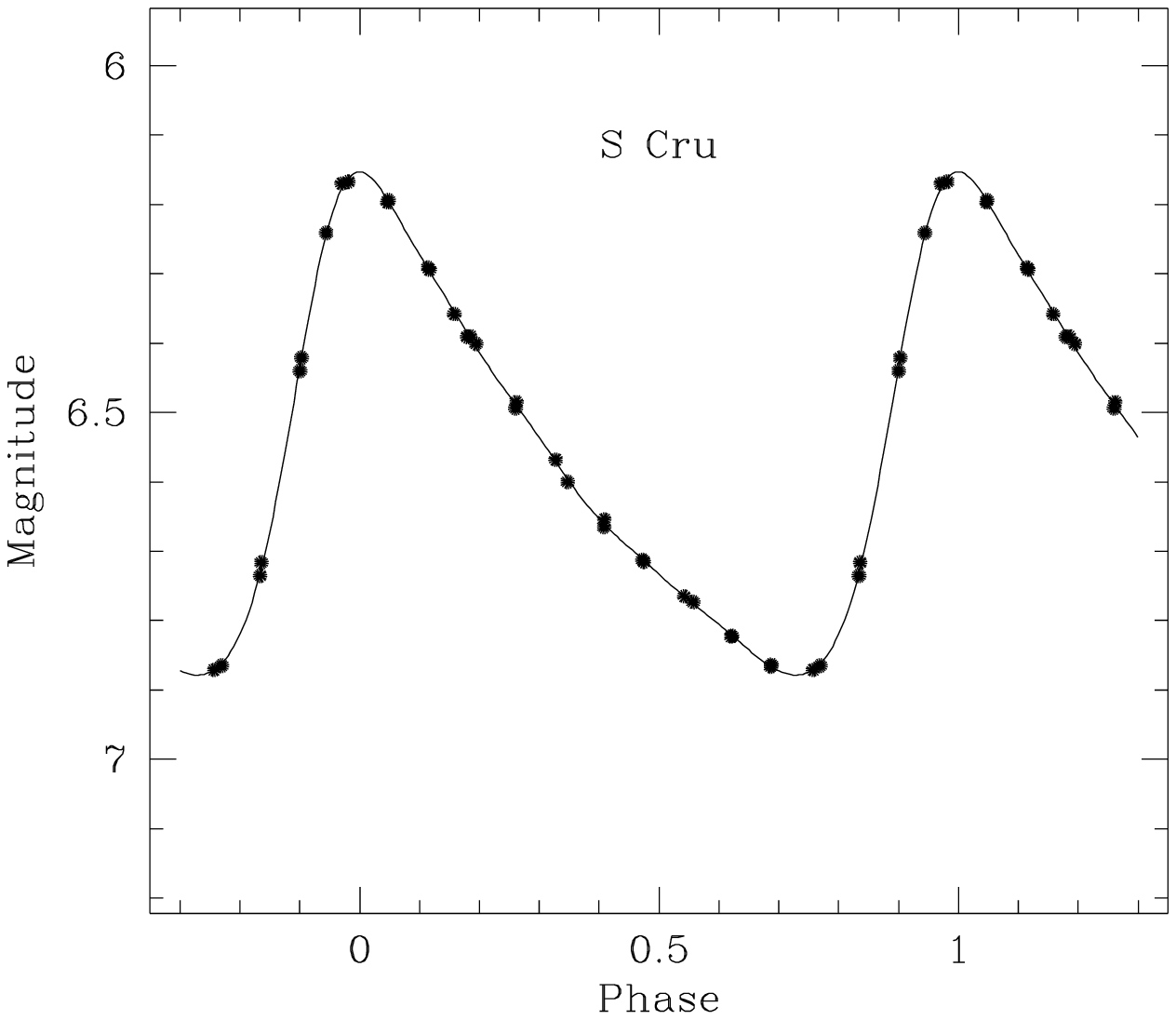,width=9truecm}}
\caption{The light curve of S Cru: a $6^{\rm th}$-order fit is necessary
to fit the steep rising branch}
\end{figure}

\section{The application to Cepheids with $P<$ 8 d}
We could verify that the sequence formed by the classical Cepheids 
is very narrow and it can be described by the linear  fit
\begin{equation}
\phi_{21}=3.332~+~0.216~P
\end{equation}
This fit is the mathematical representation  of the well-known
Hertzsprung progression.
An observed scatter of 0.30 rad in the \fdu value puts a star well
outside of the progression. By applying the Fourier decomposition to all
the available light curves of Cepheids with $P<$8 d, we could evidence
two other sequences: an upper one with
2.0 d$< P <$ 3.5 d, \fdu$>$4.2 rad and a lower one with 3.0 d$<P<$5.5~d,
\fdu$<$ 4.0 rad (Fig. 3, first panel).
 The four panels of Fig.~3 show the successive
modifications of the \fdu--$P$ plot from the first analysis (Antonello
\& Poretti 1986) to the last one (Poretti 1994).
\begin{figure}
\centerline{\psfig{figure=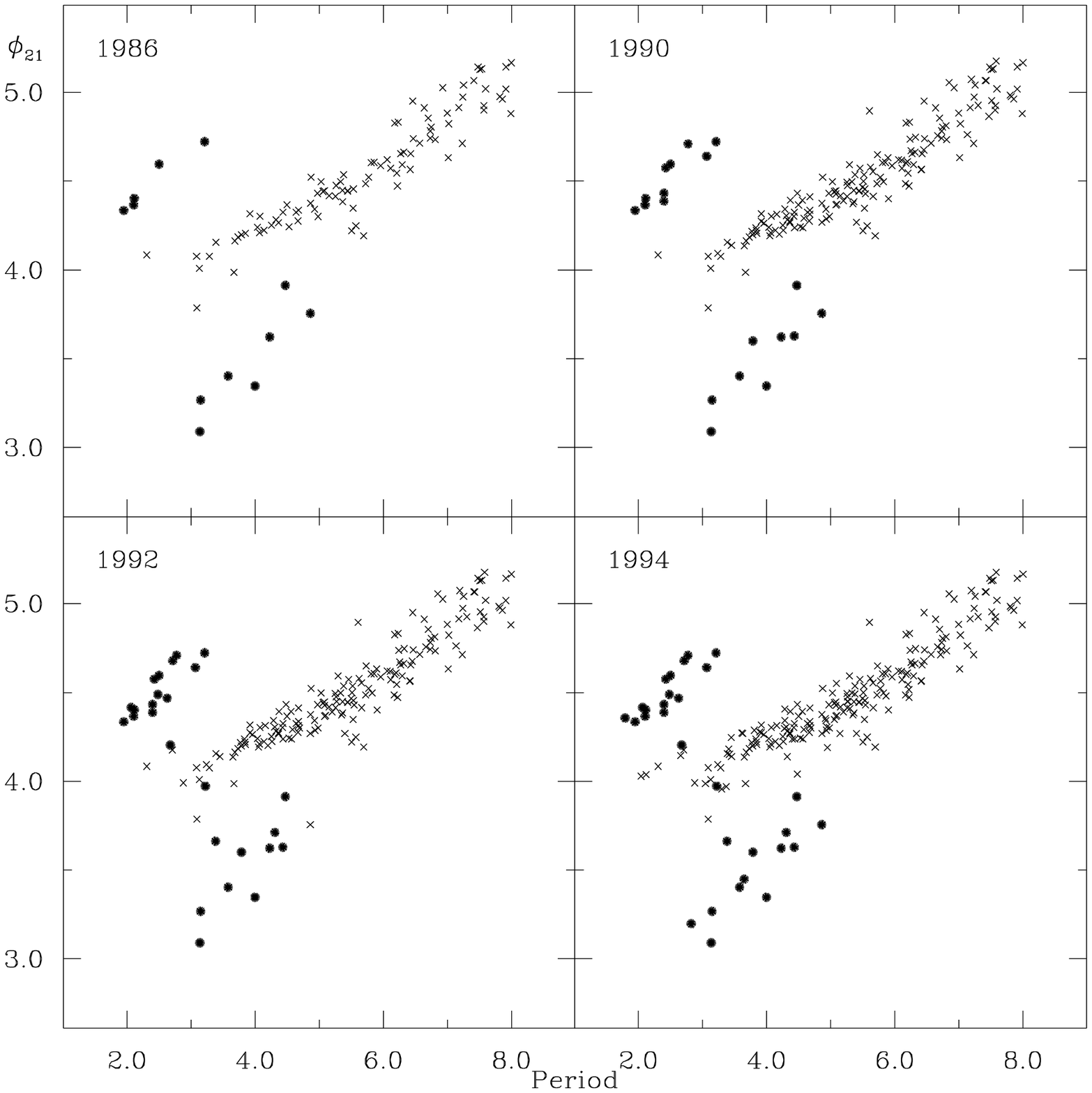,width=12truecm}}
\caption{The \fdu--$P$ plot and its modifications from 1986 to 1994. It is an
example of growing confidence in a feature, i.e. the ``Z"--shaped
progression crossing the classical, linear one.}
\end{figure}

After the first work, it was not evident what was the reason for the
three sequences.
Gieren et al. (1990)
showed that the stars located on the upper sequence are pulsating in
the first overtone mode. Different interpretations were proposed for
the stars located on the lower sequence:
Gieren et al. (1990) suggested that these stars
are fundamental pulsators but differ from classical Cepheids
for another reason, perhaps a different $M-L$ relationship.
 Antonello et al. (1990) suggested the presence of a
resonance between the first and a higher overtone at $P\approx$3 d.
They also called $C-a$ stars those forming the classical progression and
$C-b$ stars those forming the upper and lower sequences. 
Let us consider hereinafter this latter nomenclature; in Sect.~6  we
shall define the pulsational properties of the Cepheids and we shall
propose a unique classification.

The controversial aspects of the matter were an incentive to observing
a greater number of Cepheids.  The upper and lower sequences were not as
defined  as the other one; hence we decided to perform new
decompositions and, if necessary, new observations.
 For this reason our group supplied very
accurate photoelectric photometry of some selected stars whose position was
not very clear in the Fourier parameter spaces.

The new observational data collected by Mantegazza \& Poretti (1992)
brought some clarification into the matter.
 The link between the two sequences could be established by considering the
\ftu--$P$ plane. In this plane the progression is continuous and it is
formed by stars located on the upper and lower sequences in the
\fdu--$P$ plane. Hence, all the $C-b$ stars have a common nature.
But the \fdu~sequence is really interrupted by a resonance? To verify
this point, let us consider the case of BY Cas.
Its \fdu~value is quite
normal (Fig.4, left panel), as it is located on the $C-a$ sequence; however,
 its \ftu~value falls exactly on the $C-b$ sequence (Fig. 4, right panel).
It is quite evident that a star falling on the resonance interval can
display any \fdu~value: a very high one, as the stars on the upper
sequence do, a very low one, as the stars in the lower sequence do, a
``$C-a$" one, as BY Cas does. This is the signature of  a resonance.

Hence, we can summarize 6 years of investigations on the light curves of
Cepheids with $P<$ 8 d in this way:
\begin{itemize}
\item The \ftu$-P$ plane strengthens the hypothesis that $C-b$ stars (i.e.
the stars forming the upper and lower sequences in the \fdu$-P$ plane) have
a common nature; 
\item The case of BY Cas demonstrates that at $P\approx$ 3 d the
\fdu~values are spread over a wide interval. The two sequences are then
ideally connected; this is the fingerprint of  a resonance;
\item Since the theoretical models of Cepheids do not support
 resonances at $P\approx$ 3 d involving the fundamental mode,  we are forced
 to consider $C-b$ stars as first overtone pulsators;
\item the $C-a$ stars are fundamental mode pulsators, the $C-b$ stars
are first--overtone pulsators. The \fdu~values can then be successfully used
to discriminate between pulsation modes. 
\end{itemize}
\begin{figure}
\centerline{\psfig{figure=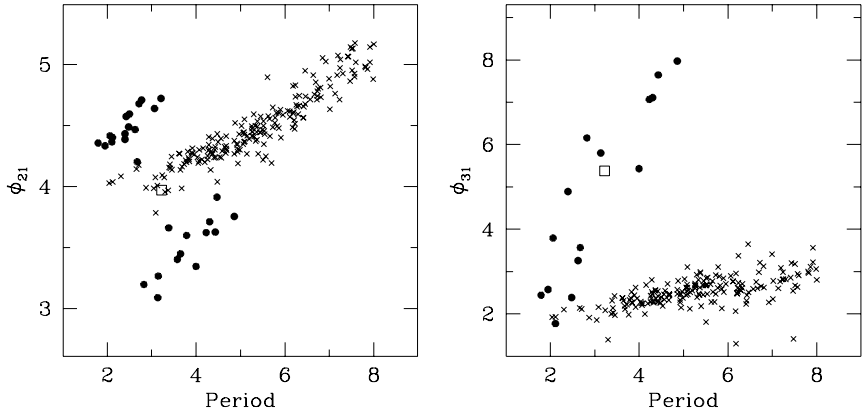,width=12truecm}}
\caption{The \ftu--$P$ plot  shows the connection between
the stars on the upper and lower sequences of the \fdu--$P$ plot. The
enlarged squares indicate the phase values of BY Cas}
\end{figure}
Hence, the suggestion firstly made by Antonello et al. (1990) 
was confirmed. The effect of the $P\approx 3$ d resonance was the same
as the one observed  for the classical Cepheid sequence at
$P\approx$10 d;  the complication here is 
the simultaneous presence of the two classes of pulsators, which
partly masks the discontinuity.

Looking at Fig.~5 we can compare the light curves of two Cepheids having similar
periods, but belonging the one to the $F$--pulsator class (BE Mon, $P$=2.705510
d) and one to the $1O$--pulsator class (V526 Mon, $P$=2.674985 d). As
can be easily noted, the light curve of BE Mon is much more asymmetric
and moreover the maximum is sharper. 

\begin{figure}
\centerline{\psfig{figure=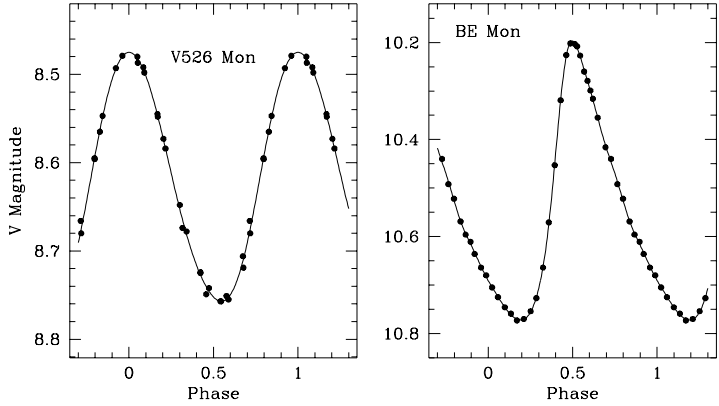,width=10truecm}}
\caption{The light curves of a Cepheid pulsating in the first overtone
(V526 Mon, left panel) is quite different from that of a Cepheid
pulsating in the fundamental one (BE Mon, right panel). The two stars
have the same period, around 2.7 d. Data were collected at ESO: the
standard deviation of the fit is 4.5 mmag for V526 Mon and 2.9 mmag for
BE Mon}
\end{figure}

\section{The double--mode Cepheids: the detection of the frequency
content}
The Double--Mode Cepheids (DMCs) supply the laboratory where the
conclusions described in the previous sections  can be verified: it is a well
established fact that in 13 cases out of 14 the two excited modes are
the fundamental and the first overtone mode. The light curve of a DMC
can be considered as the sum of the contributions of a set of
frequencies. Two are really independent (\fu and \fd); since each of the
corresponding curves is not perfectly sine--shaped, the harmonics 2\fu,
3\fu,~..., 2\fd, 3\fd,~... are also observed. Moreover, the two
independent modes are interacting and the cross coupling terms are
observed; they are defined as $|$~i\fu$\pm$j\fd~$|$ (i.e. \fm, \fp~,
2\fu+\fd, 2\fd--\fu and so on). Pardo \& Poretti (1997) submitted all the
available photometry on DMCs to a frequency analysis with the following
objectives:
\begin{itemize}
\item to quantitatively determine the importance of harmonics and of the
cross--coupling terms;
\item to compare the Fourier parameters with those of $C-a$ and $C-b$
stars;
\item to search for the fingerprints of resonances between modes, by
using Fourier parameter plots;
\item to establish properties of Fourier parameters as a function of their
order.
\end{itemize}
\begin{figure}
\centerline{\psfig{figure=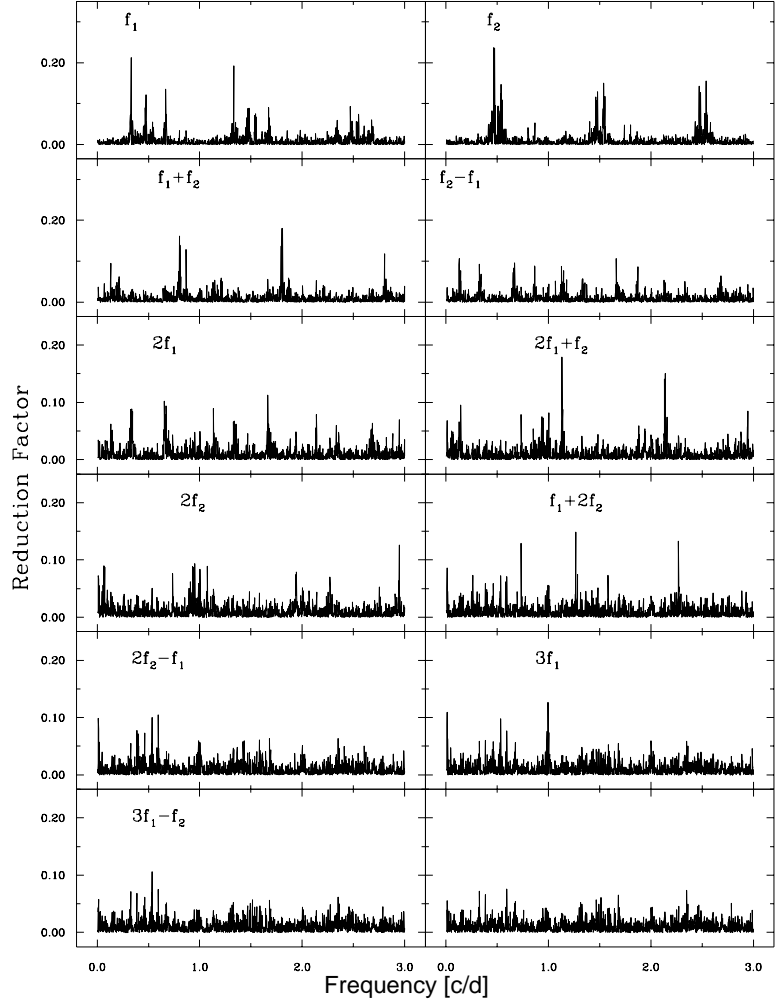,width=11truecm}}
\caption{Power spectra of the VX Pup measurements. Each panel shows the
spectrum obtained by introducing all the terms identified as known
constituents in the previous ones; this means that their frequencies are
considered as established, but their amplitude and phase values are
recalculated for each new frequency}
\end{figure}
In the approach to the light curve analysis we  took advantage of our
experience on small amplitude pulsating variables ($\delta$ Sct and
$\gamma$ Dor stars). As a matter of fact, after finding the main
constituents, the other terms have a very small amplitude and a well
tested procedure is recommended to detect them in a reliable way.
 Hence, we used the least--squares power spectrum method
(Vanicek 1971). Let us  discuss the methodology in detail using the
available measurements on VX Pup. In the first power spectrum of Fig.~6
the peak at \fu=0.3320 \cds and its alias at 1.33 \cds are clearly visible.
The aliases are particularly strong in this dataset since the
measurements were
obtained in a single site; when merging measurements obtained at two or
more
sites the height of the aliases will decrease. Then we introduced \fu~as
a known constituent (hereinafter k.c.) searching for the second term:
in the second power
spectrum the \fd=0.4674 \cd term and its whole alias structure appeared
(i.e. the 1--$f$, $f$+1, 2--$f$, $f$+2, 3$-f$  terms). 
 It is important to note that no prewhitening was done: only the frequency
value \fu~was considered as a k.c. and in the second search the unknowns were
$V_o,
A_1, \phi_1, f_2, A_2, \phi_2$. Before proceeding further with a new
frequency search,
the values of  \fu~and \fd~were refined by a simultaneous least--squares
fit and then they were introduced as k.c. in the third search, which
allowed
us to detect the \fp~term (third panel). Now, frequency refinement is a
delicate step because the third component must always satisfy the
relationship
\fp; to do this refinement, we use the MTRAP code (Carpino et al. 1987)
which
keeps this relationship locked throughout the best fit search. After the
refinement, we introduced the \fu, \fd, \fp terms as k.c.'s
($V_o, A_1, \phi_1, A_2, \phi_2, A_{f_1+f_2}, \phi_{f_1+f_2}, f_3, A_3,
\phi_3$
are the unknowns) searching for the new light curve component: we
detected
\fm. Once again, the refinement was performed by keeping the \fp~and
\fm~relationships locked; new frequency values were then obtained and
introduced as k.c.'s, the fifth component 2\fu was detected and so on.
Following this process, we detected 11 terms. We note that in  some
spectra, especially in the \fp~and 2\fu~cases, the highest peaks are not the
expected term, but their alias at 1 \cd.
This overtaking is due to the interaction between noise
and spectral window (we were dealing here with
single--site measurements).  When observing this event, the exact value
of the expected  term is considered to proceed further.
 The decision to stop the term selection
was taken when no more term was visible over the noise distribution,
i.e.
when all the terms giving a significant contribution to the light curve
shape were presumably identified. In Fig.~6 the 12$^{th}$ panel clearly shows
that no other term can be detected in a clear way as the noise distribution
is quite uniform. Of course, very small amplitude terms can remain
hidden in
the noise level, especially when dealing with inaccurate measurements.
\begin{figure}
\centerline{\psfig{figure=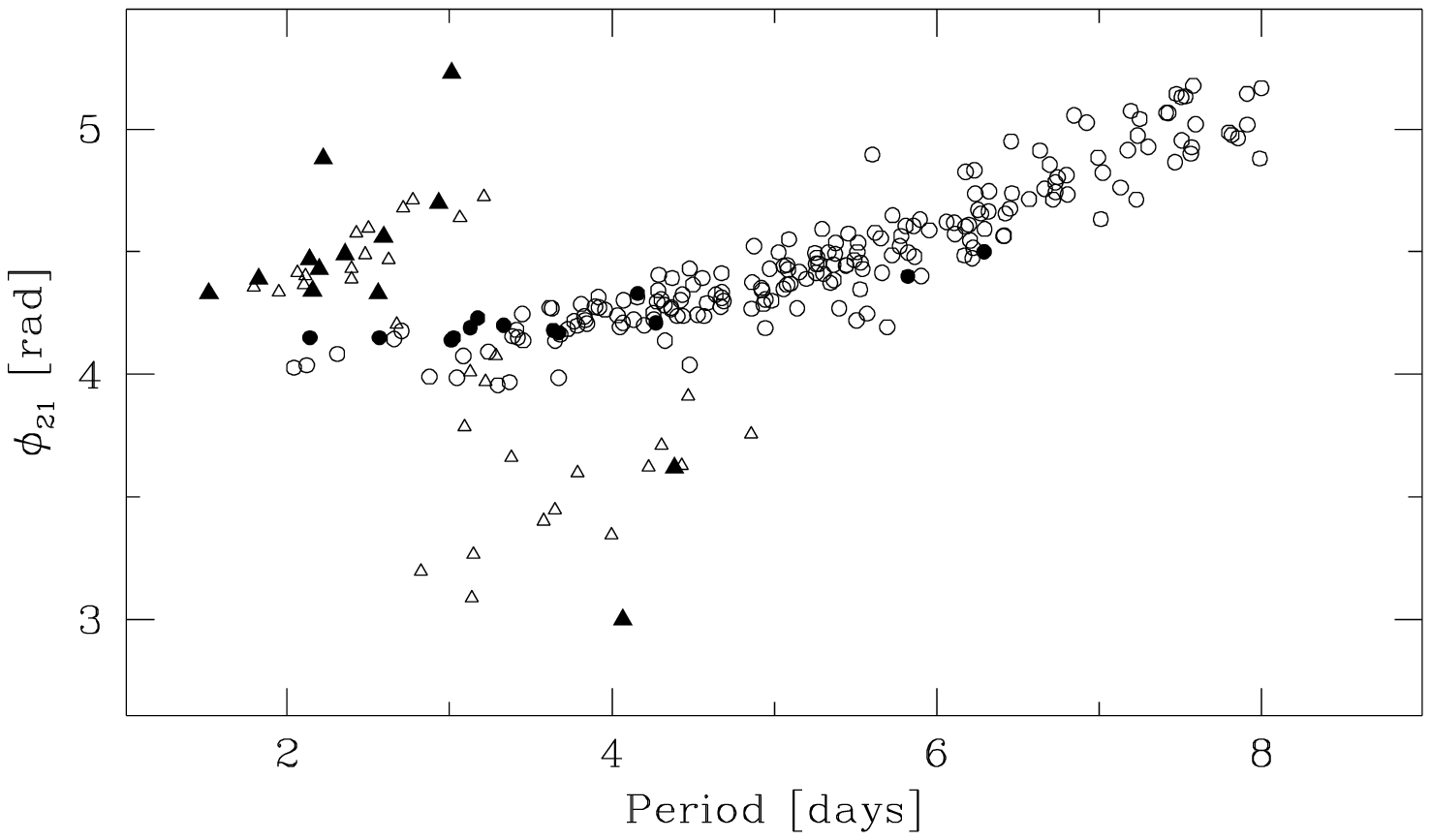,width=11truecm}}
\caption{The \fdu$-P$ plot. Dots: single mode $C-a$ Cepheids.
Triangles: single mode $C-b$ Cepheids. Filled dots: fundamental radial
modes of DMCs. Filled triangles: 1$O$ radial modes of DMCs} 
\end{figure}

\section{Comparison between double-- and single--mode Cepheids}
In the previous sections we mentioned the separation between $C-a$ and
$C-b$ stars in the space of the Fourier parameters.
 The very reliable Fourier parameters now at our disposal
for the galactic DMCs allow us to give an independent confirmation of
the proposed intrepretations. Figure 7 shows the distribution
of the $\phi_{21}$ values of
the galactic DMCs superimposed to the single--mode ones.
The $\phi_{21}$ values corresponding to the $F$ radial mode occupy
the
same region as the Classical Cepheids. In like manner, the $\phi_{21}$
values
of the the 1$O$ radial mode mimics the ``$Z$" shape: note the overlap
between
DMCs and $C-b$ in the upper part, the high value at
3.0 d (BQ Ser) and the positioning of the two $\phi_{21}$ values
belonging to
the longest period DMCs (EW and V367 Sct) just on the lower part.
It appears  quite evident that in the DMCs the light curves of the
$F$--radial
mode and the 1$O$--mode are very similar to the curves of the $C-a$
and $C-b$, respectively. In turn, this fact proves without any doubt
that $C-b$ stars are pulsating in the 1$O$ mode and that the
$\phi_{21}$
value can be considered a powerful discriminant between these modes.
It should be also noted that the $F$--mode light curve of a DMC follows the
Hertzsprung progression.
 A discontinuity is present near 3.0 d in the
light curves of 1$O$ modes of DMCs.

As a result of our step--by--step analysis, we can conclude that Cepheids can
be subdived into two groups on the basis of the different pulsation
mode:
\begin{itemize}
\item the fundamental radial mode pulsators.  They are classified
as CEP by the GCVS, are the Classical Cepheids in the current literature
and are designed as $C-a$ stars in Antonello et al. (1990);
\item the first overtone radial mode pulsators. They are
classified as DCEPS by the GCVS, are the $s$--Cepheids in the current
literature and are designed as $C-b$ stars in Antonello et al. (1990).
\end{itemize}
It should be noted that the old definition of $s$--Cepheid, i.e. a star  
showing a sinusoidal light curve, should now dropped out as too generic.
The Fourier decomposition supplies us with a quantitative tool to describe it
and small asymmetries can be measured.
As Fig. 8 shows, the stars
forming the ``$Z$" sequence also show a small $R_{21}$ value and hence
the light curve deviates very slightly from a sinewave shape.
However, it can be noted as the $R_{21}$ values for the $F$--mode
of some DMCs are smaller than the expected ones. We stressed once more
that the Fourier parameters have to be considered globally
to perform a reliable identification.

\begin{figure}
\centerline{\psfig{figure=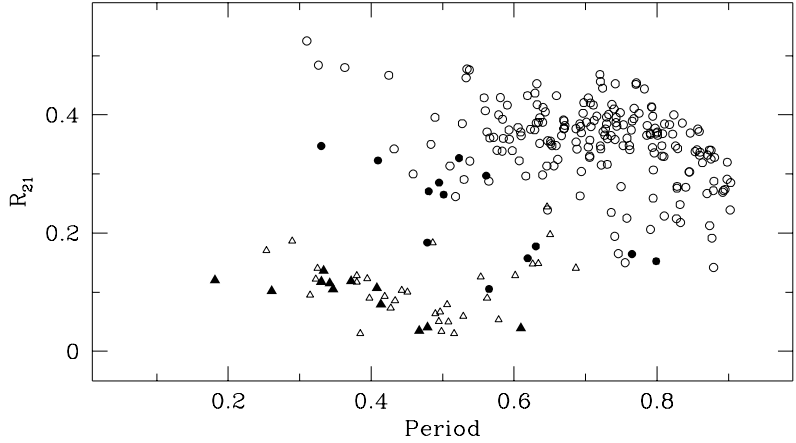,width=11truecm}}
\caption{The $R_{21}-P$ plot. Dots: single mode $C-a$ Cepheids.
Triangles: single mode $C-b$ Cepheids. Filled dots: fundamental radial
modes of DMCs. Filled triangles: 1$O$ radial modes of DMCs} 
\end{figure}

\section{The generalized phase differences}
Pardo \& Poretti (1997) fitted the $V$ magnitudes of DMC by means of the formula
\begin{equation}
V(t)= V_o + \sum_z {A_z \cos [2\pi f_z  (t-T_o) +\phi_z ]}
\end{equation}
where $f_z$ is the generic frequency, which can be an independent
frequency (\fu and \fd), a harmonic or a cross coupling term.
Their analysis demonstrated that each component
in the DMC light curves  can be defined as a combination of
two basic frequencies \fu and \fd; by defining $z=(i,j)$, we have
$f_z=f_{i,j}$=$i$\fu+$j$\fd. Some examples:
for $(i,j)$=2,0 we have the harmonic 2\fu; for $(i,j)$=1,1 the \fp~term;
for $(i,j)$=--1,1 the \fm~term; for $(i,j)$=3,--2 the 3\fu--2\fd~and so on.

In order to define the properties of the Fourier parameters of the DMC
light
curves it is very useful to recall to mind the {\it generalized phase
differences}
introduced by Antonello (1994b), here noted as $G_{i,j}$.
They are a linear combination of the phases of
each term $f_{i,j}$ and of the phases $\Phi_1$ and $\Phi_2$ of the
independent frequencies \fu and \fd.
 Their expression is given by
\begin{equation}
G_{i,j}=\phi_{i,j} - i\Phi_1 - j\Phi_2 + 2k\pi
\end{equation}

The numerical application to the U TrA fit provides some examples (the
integer $k$ values have to be selected so that $G_{i,j}\in[0,2\pi]$):
\[ G_{1,1}=\phi_{1,1} - \Phi_1 - \Phi_2 + 2k\pi = \]
\[ \hspace*{1.0truecm} = 2.93 - 5.20 - 6.25 + 4\pi = 4.04\]
\[ G_{-1,1}=\phi_{-1,1} + \Phi_1 - \Phi_2 + 2k\pi = \]
\[ \hspace*{1.0truecm} = 4.79 + 5.20 - 6.25 = 3.74 \]
\[ G_{4,1}=\phi_{4,1} - 4\Phi_1 - \Phi_2 + 2k\pi = \]
\[ \hspace*{1.0truecm} = 6.00 -4\cdot 5.20 - 6.25 + 8\pi = 4.07 \]

\begin{figure}
\centerline{\psfig{figure=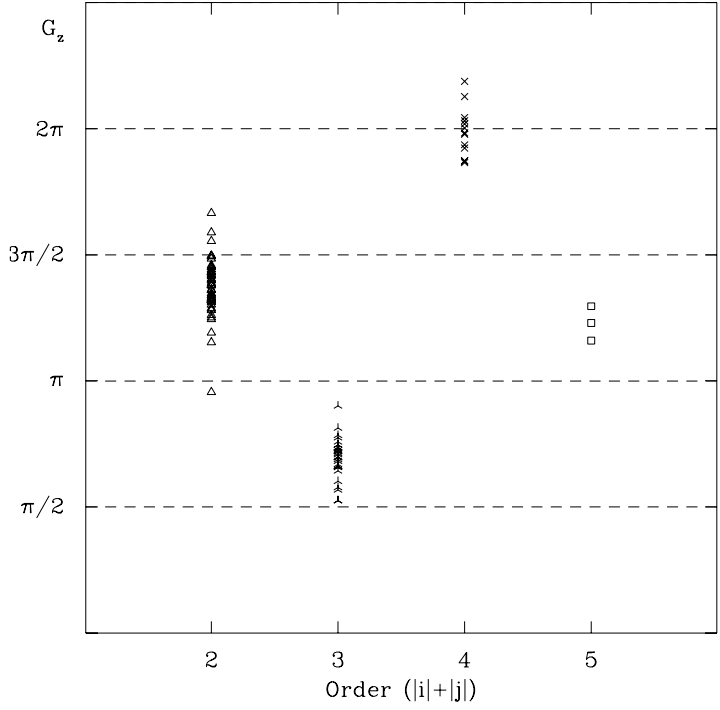,width=11truecm}}
\caption{The regularity in the Fourier parameters is emphasized when
plotting the generalized phase differences $G_{i,j}$ as function of the
order. This also means that light curves are predictable. The spread of
the 2$^{\rm nd}$ order values is discussed in the text}
\end{figure}
It is quite interesting to plot the $G_{i,j}$ values against their fit order.
Light curves of DMCs are often quoted as an example of erratic behaviour and
cycle--to--cycle variations, both in amplitude and in phase. Pardo \&
Poretti (1997) have already proved that these light curves seem to be much
 more stable than
reported and that a frequency locked fit  yields a satisfactory
representation. 

The suspicion that the DMC light curves have a
predictable behaviour is confirmed by the natural upper and lower limits
that can be easily observed in Fig. 9.
The second order terms are confined in the region just below 3/2$\pi$;
the third order terms have $\pi/2 < G_{i,j} < \pi$, the fourth order
ones
cluster around 2$\pi$ (or 0), the fifth order ones seem to have
$\pi < G_{i,j} < 3/2 \pi$.

The mean $G_{i,j}$ values are 4.30\p0.34 rad for the second order (i.e.
$\mid i\mid + \mid j\mid$=2),
2.20\p0.23 rad for the third one, 6.24\p0.31 for the fourth one, 3.85\p0.21
for the fifth one. These mean values are roughly equispaced, with a slight
tendency to increase: indeed, the differences between the mean $G_{i,j}$
of adjacent
orders are 2.10, 2.24, 2.39 rad, respectively. The latter result and the
boundary values  established above yield an experimental confirmation 
of the rule of uniformity of phase differences in monoperiodic Cepheids.
However, the observed separation ($\sim$2.2 rad) is a bit larger than expected
($\pi$/2) in the case of adiabatic pulsations in  a one--zone model (see
Poretti \& Pardo 1997 for a more detailed discussion).

\begin{figure}
\centerline{\psfig{figure=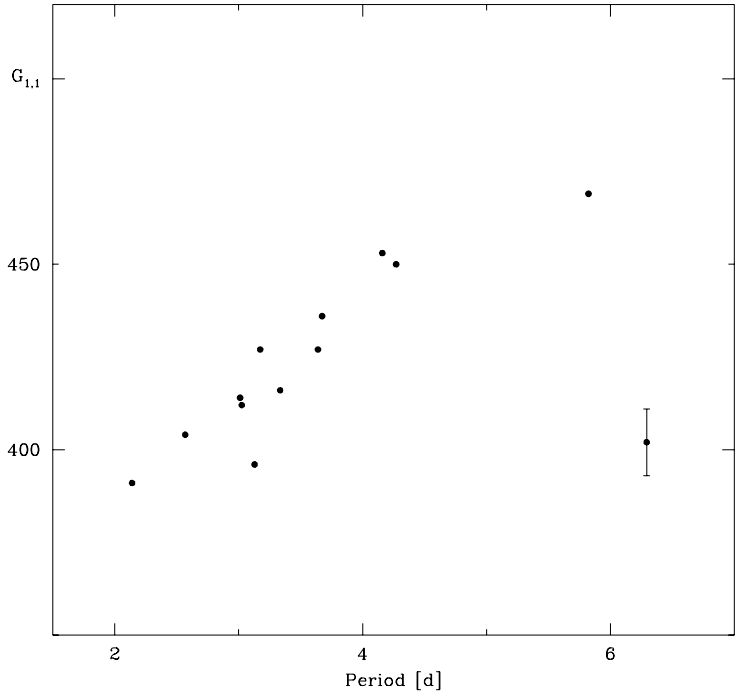,width=11truecm}}
\caption{The $G_{1,1}$ progression related to the coupling term \fu+\fd.
Note how the last point (related to V367 Sct) appears to deviate
strongly. This can be explained by a resonance acting at $P\approx$6.0 d}
\end{figure}

The second order $G_{i,j}$ values (2\fu, \fp, \fm, 2\fd) range from
3.00 to 5.23 rad; it was expected to see a little
spread of the $G_{0,2}$ values owing to the resonance at 3.0 d.
Indeed the two extrema are just related to
the 2\fd~components  of the BQ Ser (the DMC approaching resonance from
the shorter periods) and EW Sct (the DMC approaching resonance from
the longer periods) light curves. Antonello (1994a) reported
another possible resonance between the third overtone and the \fp~term
near 6.5 d; Fig.~10 definitely proves it.  
The last point (4.02\p0.09 rad, V367 Sct, $P$=6.293 d) is
clearly out of the  progression followed by the other points. 
Combined with the progressive weakening of the amplitude of the \fp~term,
this fact  strongly supports the action of a
resonance effect involving the \fp~term.

\section{The resonance signatures and the influence on the models}

There are many effects ascribed to resonances between modes:
\begin{enumerate}
\item The evidence of the resonance at $P\approx10$ d was firstly
evidenced by Simon \& Lee (1981). The values of the \fdu~parameter were
spread on a very large interval and the progression is abruptly
interrupted. The involved modes are the second overtone and the
fundamental mode ($2O/F$=2);
\item In this paper we reconstructed the methodological procedure used to
show how we recognized the effect of the resonance at $P\approx$3 d in 
the 1$O$ pulsator light curves; the involved modes are the fourth and
the first overtone. Aikawa (1993) tried to obtain the first
implications of its effect in nonlinear models;
\item A resonance is expected around 6--7 d for fundamental pulsators,
involving the fourth overtone and the fundamental modes (Moskalik et al
1989). The small feature related to it was noted by Antonello (1994a);
\item As regards the DMCs, the Fourier decomposition of the light curve of
V367 Sct suggests a possible resonance around 6.5 d. In such a case, the
cross--coupling term \fu+\fd~and the third overtone were involved;
\item When considering longer periods, features suggesting the action
of two resonances were observed (Antonello \& Morelli 1997).
The fundamental and the third overtone are the involved modes for
that at $P\approx$ 27 d ($3O/F$=3), while the fundamental and the first overtone
are those for that at $P\approx$ 24 d ($1O/F$= 3/2)
\end{enumerate}
It should be noted that their effects are not very strong and a careful
analysis is necessary to identify them. Moreover, other dedicated,
photometric observations can be useful.
For an application of the same technique to radial
velocity curves and related results see Kienzle et al. (1999).
\begin{figure}
\centerline{\psfig{figure=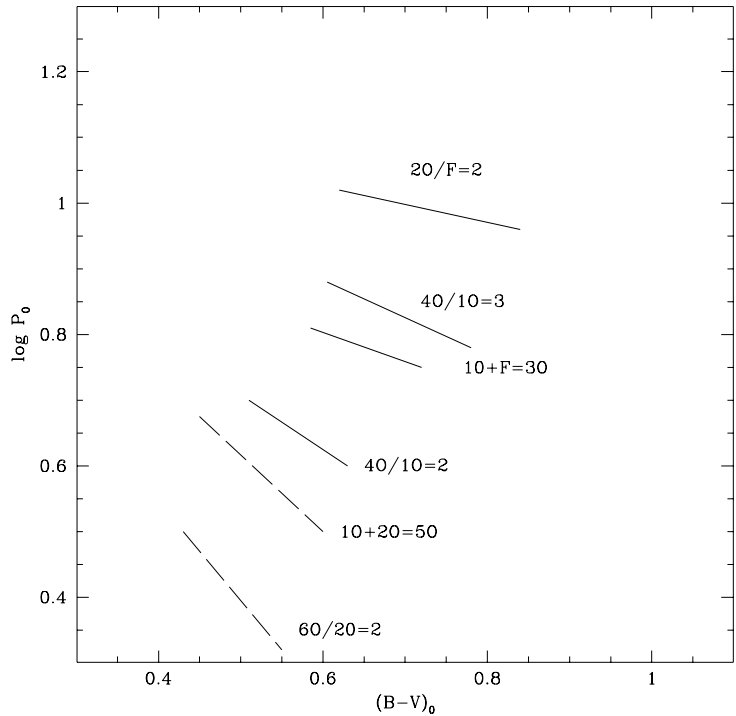,width=11truecm}}
\caption{The resonances predicted by the linear model combined with
nonstandard $M-L$ relationship are shown. $F$ indicates the fundamental
radial mode; 1$O$, 2$O$, 3$O$ the first, second, third... radial
overtones. Dashed lines indicated resonances not (yet?) observed}
\end{figure}
Figure 11 shows the positions of the resonances and the involved modes in a
$(B-V)-P$ plot. They were predicted by linear models and for nonstandard
Mass--Luminosity relationship. Theoretical models can be obtained by
varying some input parameters as opacity, overshooting effects,
$M-L$ relationship. A general agreement between theoretical
models and observational effects was found. Sequences of models were
realized using both standard and overshooting--type $M-L$ relationships
(Antonello 1997). Other models were obtained by using different
artificial viscosity parameters and temperature values. In all these
cases the theoretical light curves were decomposed and the Fourier
parameters straightly compared with the observed ones.
Recent results indicated  that the models with mild overshooting
have $M-L$ relationships in better agreement with observations than others
based on standard assumptions (Antonello 1997).
 
\section{Conclusion}
This  paper shows how the analysis of the light curves can be used to
probe the structure of the Cepheids.  Therefore, we can study the Cepheids
from the point of view of

\vskip 0.3truecm
\centerline{\it asteroseismology}
\parindent=0truecm
\vskip 0.3truecm

since  to find resonance effects is as to sound stellar interiors.
The discontinuities at $P\approx$10 d and $P\approx$3 d were confirmed by the
data obtained in the framework of large--scale projects as MACHO and EROS 
(observations of Cepheids located in the Small and Large Magellanic
Clouds). It will be interesting to carefully check also the other in the
same large databases.

\end{document}